\documentclass[journal]{article}
\usepackage{amsfonts,amssymb,anysize}
\usepackage{times,mathptm}
\usepackage{amsmath,amsthm,graphicx}
\usepackage{hyperref}
\marginsize{0.9in}{0.9in}{0.6in}{0.8in}

\newtheorem{definition}{Definition}
\newtheorem{example}{Example}
\newtheorem{theorem}{Theorem}

\begin{document}

\title{Quantum Circuit Placement\thanks{
Copyright $\copyright$ IEEE.   Reprinted from IEEE Transactions on 
Computer-Aided Design of Integrated Circuits and Systems, 27(4):752--763, April 2008. \newline
This material is posted here with permission of the IEEE.  Internal or
personal use of this material is permitted.  However, permission to
reprint/republish this material for advertising or promotional purposes or
for creating new collective works for resale or redistribution must be
obtained from the IEEE by writing to
\href{mailto:pubs-permissions@ieee.org}{pubs-permissions@ieee.org}. \newline
By choosing to view this document, you agree to all provisions of the
copyright laws protecting it.}}

\author{Dmitri~Maslov\thanks{D. Maslov is with the Institute for Quantum Computing, University of Waterloo,
Waterloo, ON, N2L 3G1, Canada, email: dmitri.maslov@gmail.com.},
		Sean~M.~Falconer\thanks{S. M. Falconer is with the Department of Computer Science, University of Victoria,
Victoria, BC, V8W 3P6, Canada.},
		Michele~Mosca\thanks{M. Mosca is with the Institute for Quantum Computing, University of Waterloo, and
Perimeter Institute for Theoretical Physics, Waterloo, ON, Canada.}
}

\maketitle

\begin{abstract}
We study the problem of the practical realization of an abstract
quantum circuit when executed on quantum hardware.  By practical, we
mean adapting the circuit to particulars of the physical environment
which restricts/complicates the establishment of certain direct
interactions between qubits.  This is a quantum version of the
classical circuit placement problem.  We study the theoretical
aspects of the problem and also present empirical results that match
the best known solutions that have been developed by
experimentalists.
Finally, we discuss the efficiency of the approach and scalability of
its implementation with regards to the future development of quantum
hardware.
\end{abstract}

\section{Introduction}
Quantum algorithm design is usually done in an abstract model of
computation with the primary interest being the development of {\it
efficient} algorithms (e.g. polynomial versus exponential sized
circuits) independently of the details of the potential physical
realization of the related quantum circuits.
Theoretical quantum computational architectures usually, for
convenience, assume that it is possible to directly interact
any two physical qubits.
It is well known that in physical implementations\footnote{
 Trapped ions \cite{ar:s-k} and
liquid state NMR (Nuclear Magnetic Resonance) \cite{tr:clkv} are two of the most
developed quantum technologies targeted for computation (as opposed to communication
or cryptography). Other state of the art quantum information processing
proposals are described in \cite{www:h}.} direct
interactions between two distinct qubits may sometimes be hard if
not impossible to establish.
For instance, it is not uncommon for an
interaction to have less than $0.2$ Hz frequency (e.g. carbons
C2 and C6 from \cite{ar:nmrd}), while the decoherence may take
around one second. This means that the interaction between such
qubits will be essentially seen as noise. A physical realization of
a quantum processor can of course {\it indirectly} couple any two
qubits, with some overhead cost. However, in practice this can be
very costly (and, for example, make the implementation infeasible
with available technology), and so one needs to find a way to
minimize this overhead, which we will refer to as quantum circuit
placement problem.

Some authors \cite{ar:mi} try to avoid long interactions by
considering the linear nearest neighbor computational architecture where all
qubits are lined up in a chain (such as trapped ions
\cite{ar:s-k}) and only the nearest neighbors are
allowed to interact. Mathematically, for a system with $n$ qubits
$q_1,\;q_2,...\;,q_n$, the 2-qubit gates are allowed on qubits whose
subscript values differ by one. Such an architecture, however,
{\em does not necessarily} represent reality.  For instance, in NMR
technologies the fastest interactions
will not usually form a chain. In a 3-qubit computation, the fastest
interactions can be aligned to form a chain (acetyl chloride, Figure
\ref{ac}, \cite{martin}) or a complete graph \cite{ar:cjfs}.  In all
molecules used in NMR and known to the authors that employ four or
more qubits, the fastest interactions are established along the
chemical bonds and they do not form a single chain.  The chain
nearest neighbor architecture can apply to some quantum dot
based architectures and certain variations of trapped ion \cite{ar:s-k} quantum
technologies, but does not represent the reality behind
other technologies, including those based on NMR, Josephson
junctions, optical technology, and certain other variations of
trapped ions \cite{ar:mo}. Furthermore, we are unaware of any
automated procedure for optimizing the interactions even if the
latter are restricted to the chain nearest neighbor architecture.

It has been recently proposed to use EPR pairs to teleport
logic qubits in a Kane model \cite{ar:sdk} when long-distance swapping
of a {\em single} value becomes too expensive \cite{ar:coi}.
However, this type of ``technology mating'' may not always
be possible. For instance, it seems hard if not impossible to mate
NMR with an optic EPR source while doing a computation on an NMR machine.
It is also not obvious how congested such an architecture gets if simultaneous
teleporting of a large number of qubits is required.
We concentrate on an approach that does not require mating
different technologies, and is {\em technology independent}.

At roughly the same time as our initial publication\footnote{This paper is an
extended version of our earlier conference publication presented at DAC-2007.} there was another publication devoted
to the study of quantum circuit placement problem \cite{arXiv:0704.0268v1}.
Both papers, \cite{arXiv:0704.0268v1} and ours consider the problem
of placement in the quantum circuit computational model. The major
difference that sets the two research directions apart and yet shows
that the results are complementary is in our paper we consider pre-given
architectures, whereas \cite{arXiv:0704.0268v1} forms an architecture
on the fly according to the circuit that needs to be implemented.

Unlike in the quantum case, the classical circuit placement
problem is very well known and extensively studied \cite{bk:lms}.
Classical circuit placement is concerned primarily with minimizing
the total wirelength, power, and congestion. Timing minimization is
considered to be a secondary optimization objective. In quantum
technologies, placement is equivalent to timing optimization under
the natural assumption that gate fidelities \cite{bk:nc} are
inversely proportional to the coupling strength/gate runtime,
otherwise, a function of both may be considered. However, timing
minimization for quantum versus CMOS technologies are two distinct
and dissimilar problems. This is not surprising given the conceptual
differences in the two computational models.
A quantum circuit placer cannot be constructed by reusing the
existing EDA tools and thus, must be built from scratch.

In practice and at present, the mapping of qubits in a circuit into
physical qubits is done by hand, but as the quantum computations
scale, a more robust (automated) technique for such a mapping must
be employed. In this paper, we study the problem of mapping a
circuit into a physical experiment where we wish to optimize the interactions.
We formulate this problem mathematically and study its complexity.
Given the problem of finding an optimal assignment of qubits is
unlikely to have a polynomial time optimal solution, as it is
NP-Complete, and exhaustive search requires $n!$ tries for an
$n$-qubit system, we develop a heuristic-based algorithm. We verify the latter
on data taken from existing quantum experiments.

Throughout the text, and especially in the experimental results
section, we refer to liquid state NMR technology for quantum
computation. While there are certain problems with using liquid
state NMR to work with over 10 qubits \cite{tr:clkv, ar:j, ar:nmrd}, so
far, it is the best developed methodology for quantum computing
experiments. Liquid state NMR experiments provide a nice data set
for both the interaction frequencies and the circuits of interest,
and a test bed for development of suitable scalable spin-based
quantum computing devices. While we use results from the liquid
state NMR experiments, it is important to emphasize that our approach
is not restricted to a particular technology. For instance, many
circuits we consider are composed for the Ising type Hamiltonian,
however, Ising type Hamiltonian describes not only the liquid state NMR experiments,
but also certain superconductivity proposals \cite{quant-ph/0603224}.

The remainder of the paper is organized as follows. We start with a basic
overview of quantum computation and description of quantum circuits, in
Section \ref{sec:p}.  Following this, in Section \ref{sec:f}, we
formally define the objects we will
be working with and formulate the circuit placement problem. We next
study the complexity aspect of the circuit placement problem, and conclude
that even a simplified version of this problem is NP-Complete, refer to
Section \ref{sec:c}. Thus, for the purpose of having an efficient practical
solution to the placement problem we need to concentrate on heuristic
solutions. Our heuristics, described in Section \ref{sec:hs}, consist
of two algorithms: determining an efficient placement of individual
subcircuits, and
permuting the qubit values to interconnect placed subcircuits.
We conclude this section with the description of our quantum circuit placement
software package. In Section \ref{sec:er} we describe a number
of experiments that test quality, efficiency, and scalability of our
heuristic solution. Concluding remarks can be found in Section
\ref{sec:conclusions}.

\section{Preliminaries}\label{sec:p}

We present a short review of the basic concepts of quantum
computation. For a more detailed introduction, please see \cite{bk:nc}.

The key ingredients of the standard quantum computation model
are the data structure (described by state vectors),
the set of possible transformations allowing one to perform a
desired calculation (unitary matrices),
and the access to the result of a computation (measurement).
In very basic terms, every quantum algorithm begins with a state initialization,
followed by a computational stage, and ends with a measurement.
The description of the state vectors, unitary matrices, and
quantum measurement is the focus of the following discussion.

The state of a single qubit is a linear combination $\alpha_0
|0\rangle + \alpha_1 |1\rangle$ (which can also be written as $\sum_{x \in \{0,1\}}\alpha_x|x\rangle$)
in the basis $\{|0\rangle,\;|1\rangle\}$, called the {\em state vector}, where $\alpha_0$
and $\alpha_1$ are complex numbers such that
$|\alpha_0|^2+|\alpha_1|^2=1$. The real numbers $|\alpha_0|^2$ and
$|\alpha_1|^2$ represent the probabilities of reading the
``pure'' logic states $|0\rangle$ and $|1\rangle$ upon {\em measurement}.
The state of a quantum system with $n>1$ qubits is given by an
element of the tensor product of the single state spaces and can be
represented by a normalized vector of length $2^n$ ({\em state vector}),
as $\sum_{x \in \{0,1\}^n}\alpha_x|x\rangle$,
where $\sum_{x \in \{0,1\}^n}|\alpha_x|^2=1$.
Each of the real numbers $|\alpha_x|^2$ represents the probability of
finding the quantum mechanical system in the state $|x\rangle$.
As such, measurement gives probabilistic access to $n$ bits of classical information.
According to the laws of quantum mechanics, a quantum system evolution allows changes
of the state vector through its multiplication by $2^n \times 2^n$ {\em unitary matrices}
(a square matrix $U$ over complex numbers is called unitary iff $UU^{\dagger}=I$,
where $U^{\dagger}$ is the complex conjugate of $U$).

For example, a
2-qubit quantum mechanical system described by the state vector $\vec{v}=|10\rangle$
(the result of the measurement of this state is deterministic, since there is
only one non-zero coordinate) may be acted on by a unitary
matrix of appropriate dimension (in this case, $2^n|_{n=2}=2^2=4$) 
\begin{equation}\label{mat:u}
U=\frac{1}{2}\left( \begin{array}{rrrr}
1 & 1 & 1 & 1 \\
1 & i & -1 & -i \\
1 & -1 & 1 & -1 \\
1 & -i & -1 & i
\end{array} \right).
\end{equation}

\noindent This results in the transformation $\vec{v} \mapsto U\vec{v}$
computed as follows:

\noindent $\vec{v}=|10\rangle=0*|00\rangle+0*|01\rangle+1*|10\rangle+0*|11\rangle=\left( \begin{array}{r}
0\\
0\\
1\\
0
\end{array} \right) \mapsto
$ $ U\vec{v} = \frac{1}{2}\left( \begin{array}{rrrr}
1 & 1 & 1 & 1 \\
1 & i & -1 & -i \\
1 & -1 & 1 & -1 \\
1 & -i & -1 & i
\end{array} \right)\left( \begin{array}{r}
0\\
0\\
1\\
0
\end{array} \right)$ \newline \noindent$= \frac{1}{2}\left( \begin{array}{r}
1\\
-1\\
1\\
-1
\end{array} \right) = \frac{1}{2}(|00\rangle-|01\rangle+|10\rangle-|11\rangle).$

\noindent With this state, the result of measurement gives $|00\rangle$, $|01\rangle$, $|10\rangle$ and $|11\rangle$
with equal probabilities. In fact, with this example we have illustrated application of the
2-qubit Quantum Fourier Transform to the basis state $|10\rangle$.

The above shows how a state vector evolves under the action of a unitary transformation. 
In practice, a desired unitary transformation will need to be decomposed into primitive 
operations that can be implemented in hardware.  At the experimental level, these primitive 
operations, or elementary gates, are usually defined in terms of the Hamiltonian (which 
describes the energy of a quantum mechanical system), 
and depends on a particular implementation/technology.  While our
placement algorithms/theory and program implementation {\em readily support any
gate library}, we next introduce some of the particular gates used in
the examples included in this paper (in fact, the following family of gates
forms a complete library).


\begin{itemize}
\item single qubit rotation gates
$R_x(\theta):=\left(\cos(\frac{\theta}{2}) \;\;\;\;\; -i\sin(\frac{\theta}{2})
\atop -i\sin(\frac{\theta}{2}) \;\;\;\;\, \cos(\frac{\theta}{2})\right)$,
$R_y(\theta):=\left(\cos(\frac{\theta}{2}) \;\;\;\; -\sin(\frac{\theta}{2})
\atop \sin(\frac{\theta}{2}) \;\;\;\;\;\;\; \cos(\frac{\theta}{2})\right)$,
and $R_z(\theta):=\left( e^{-i\theta/2} \;\;\;\;\;\; 0
\atop 0 \;\;\;\;\;\;\;\;\; e^{i\theta/2}\right)$;

\vspace{1mm}
\item two-qubit interaction gate \newline
$ZZ(\theta):=\left( \begin{array}{rrrr}
e^{-i\theta/2} & 0 & 0 & 0 \\
0 & e^{i\theta/2} & 0 & 0 \\
0 & 0 & e^{i\theta/2} & 0 \\
0 & 0 & 0 & e^{-i\theta/2}
\end{array} \right)$.

\end{itemize}

In liquid state NMR, single qubit $R_x$ and $R_y$ gates are implemented by radio frequency (RF) pulses in the
$X$-$Y$ plane (in the lab setting this usually corresponds to the horizontal plane of the physical 3D space
we live in).  The $R_z$ gates are implemented by a change of the so-called rotating reference frame
and require no action/waiting.  From the designer's perspective this means that the $R_z$ gates are
``free''. Finally, $ZZ$ gates are a part of the drift Hamiltonian, meaning they
happen as a natural evolution of the system and must be waited for until completion.
Those $ZZ$ interactions/gates that are not needed in a computation get eliminated via a technique
called refocussing.

In particular, in terms of the NMR gates, the matrix in equation (\ref{mat:u}) can be decomposed into the
matrix product
\begin{eqnarray*}
I\otimes R_x(\pi)\;\cdot\; I\otimes R_z(\pi)\;\cdot\; I\otimes R_y(-\frac{\pi}{2}) \;\cdot\;
R_z(-\frac{3\pi}{4})\otimes R_z(-\frac{3\pi}{4}) \\
\cdot\; ZZ(-\frac{\pi}{4}) \;\cdot\; R_y(\frac{\pi}{2})\otimes I \;\cdot\; R_z(-\pi)\otimes I \;\cdot\; R_x(\pi)\otimes I
\end{eqnarray*}

\noindent up to the global phase (i.e., constant factor) $e^{i\pi/8}$ 
and a permutation of the output qubits, both of which are insignificant
considerations from the point of view of the computation that needed to be performed. Note, that the
circuit associated with the above implementation requires only four pulses and uses only one two-qubit gate.
This circuit representation was specifically optimized for the liquid state NMR implementation.

Liquid state NMR works in a weak interaction mode, meaning the two-qubit $ZZ$ gate takes
much longer (e.g., often on the order of 10 times longer) to be applied than a
single qubit gate. Thus, careful assignment of the interactions used in the computation
is crucial in terms of the efficient execution of a logical quantum circuit on a physical
quantum mechanical system; this is the topic of study addressed in the present paper.

Finally, note that $ZZ(\pi/2)$ is equivalent to CNOT (a commonly used gate that transforms basis states
according to the formula $|a\;\;b\rangle \mapsto |a\;\;a \oplus b\rangle$, and extends to other
quantum states by linearity) up to
single qubit rotations. This means that every circuit with single qubit and CNOT gates
can be easily rewritten in terms of single qubit rotations $R_x$, $R_y$ and $R_z$, and the 
$ZZ(\pi/2)$ gates, and such a rewriting
operation does not change a particular instance of the associated placement problem.

\section{Formalization of the circuit mapping problem}\label{sec:f}

Circuits devised by researchers in the area of quantum computing are very
often written in terms of a single qubit and two-qubit gates.
Thus, any circuit with gates spanning over more than two qubits is
first translated into a circuit with a single qubit and two qubit
gates. We assume such a circuit as input. Next, most of the practical circuits
are levelled, that is the gates that can be applied in parallel appear at
one logic level (in circuit diagrams: directly below or above other gates from
the same level). Levelization helps to reduce the overall runtime of
the circuit, and thus it is a desired operation at the time the circuit
is synthesized.

Before the circuit's qubits (i.e. logical qubits) are mapped into the
molecule's nuclei (i.e. physical qubits) and the exact delays are known,
it is hard to decide how long it will take to execute
a single gate and take it into account while trying to minimize
the delay of the entire circuit. This means that the
circuit's qubits must be mapped to the molecule's nuclei first.

In a practical setting and according to how
it is done in current experiments the {\em first} step
is to efficiently assign qubits to nuclei. In the existing NMR tools,
the timing optimization is built into a compiler \cite{ar:bjkl} that takes in a circuit
and a refocusing scheme and outputs a sequence of (timed) pulses ready to be
executed. This is the last step before the circuit gets executed and before it is
done the logical qubits must be assigned to the nuclei.

In practice, it is a natural assumption that gates from the next
level can start being executed before execution of the current level
has completed. The total {\bf runtime} is defined as the time
spent by a circuit between the input initialization and measurement
of the output. With the assumptions made, the runtime of a circuit composed
of gates $G[1],G[2],...,G[k]$ can be found by the following dynamic
programming algorithm.

{\footnotesize
\begin{verbatim}
Create array Time[1..n], initiate its elements to 0;
for i=1..k
  if G[i] is a 2-qubit gate built on qubits t and c
    time[c]:=
       max{time[c],time[t]}+GateOperatingTime(G[i]);
    time[t]:=time[c];
  else // else gate G[i] is a single qubit gate that operates on qubit t
    time[t]:=time[t]+GateOperatingTime(G[i]);
  end if;
end for;
return max{time[1..n]}; // maximally busy qubit is the one that finishes the job last
\end{verbatim}
}

Circuit runtime calculation in a computational model where logic levels
need to be executed sequentially is also supported by the theory we develop
and our program implementation. We next introduce the following
notations/definitions.

\begin{definition}
A {\bf physical environment (molecule)} is a complete non-oriented graph
with a finite set of vertices (nuclei) $\{v_1,v_2,$ $...\;,v_m\}$ and weighted edges
$\{(v_i,v_j)\}_{1 \leq i \leq j \leq m}$ with $W(v_i,v_j)\geq 0$.
\end{definition}

Weights $W(v_i,v_j)$ are proportional to the inverse of the coupling
frequency (for two qubit gates) and the inverse of nuclei processing
frequencies (for single qubit gates).  In other words, they indicate
how long it takes to apply a fixed angle 2-qubit gate (in case of
edge $(v_i,v_j)$, $i \neq j$) or a fixed angle single qubit gate (in
case of edge $(v_i,v_i)$).

\begin{definition}
A {\bf quantum circuit} is built on $n$ qubits $q_1,q_2,...$ $,q_n$
and consists of a finite number of levels $L_1,L_2,...\;,L_k$,
where each level $L_i$ is a set of one or two qubit gates $G_i^1,G_i^2,...,G_i^{t_i}$.
Each gate has a weight function $T(G_i^j)$ that indicates how long it takes
this gate to use the interaction defined by the qubits it operates on
\footnote{For example, $T(R_x(180))=2*T(R_x(90))$ because it takes twice
the time to do a 180-degree rotation as compared to a 90-degree rotation.}.
\end{definition}

\begin{definition}
The {\bf quantum circuit placement problem} is to construct an injective (one-to-one)
function $P:\{q_1,$ $q_2,...\;,q_n\} \mapsto \{v_1,v_2,...\;,v_m\}$ such that with
this mapping the runtime of a given circuit is minimized.
The gate's $G(q_i,q_j)$ execution cost is defined by the mapping $q_i\mapsto v_i$,
$q_j\mapsto v_j$ according to the formula 
$$GateOperatingTime(G(q_i,q_j)):=W(v_i,v_j)*T(G(q_i,q_j)).$$
\end{definition}

There are $\frac{\displaystyle m!}{\displaystyle (m-n)!}$ potential
candidates for an optimal matching. Thus, a simple minded exhaustive search
procedure will require runtime at least $O(\frac{\displaystyle m!}{\displaystyle (m-n)!})$,
counting a singe operation per each possible assignment.

\begin{example}

Illustrated in Figure \ref{ac}(a) is the acetyl chloride molecule and a
table with the chemical shifts/coupling strengths (in liquid state
NMR) \cite{martin}. The chemical shifts and coupling strengths were
recalculated into the delays needed to apply 90-degree rotations in
the $X$-$Y$ plane (single qubit rotations around the axis $Z$ are
``free'' in that they are done by changing the rotating reference
frame \cite{ar:klmt} and require no pulse or a delay associated with
respect to their application \cite{tr:clkv}) and a 90-degree
$zz$-rotation (2 qubit gate). The delays are measured in terms of
$\frac{1}{10000}\; sec$, and are rounded to keep the numbers
integer. Figure \ref{ac}(b) describes liquid state NMR computational
properties of the molecule in graph form.

\begin{figure}[ht]
\begin{center}
\includegraphics[height=42mm]{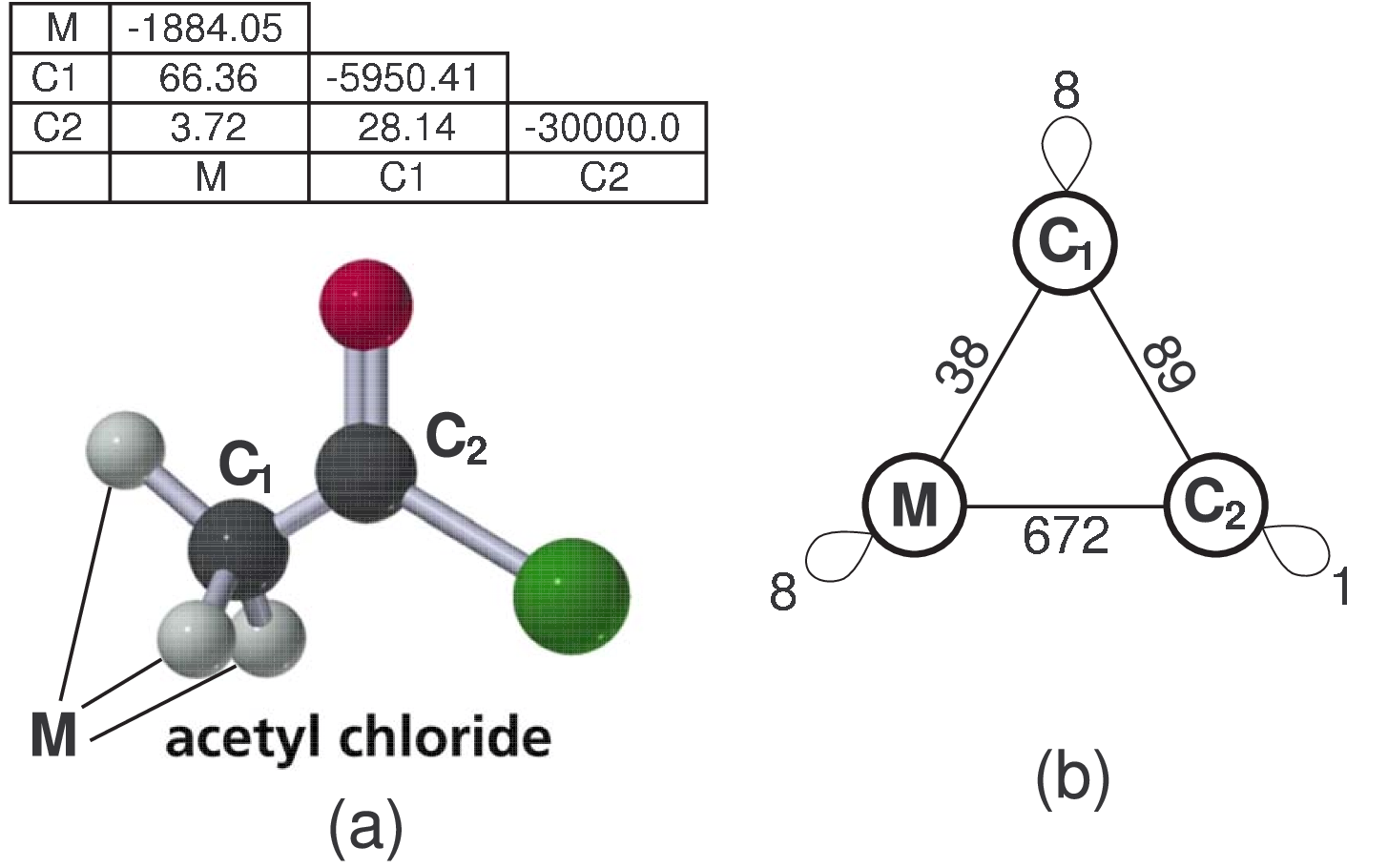}
\caption{(a) acetyl chloride molecule with three spin-$\frac{1}{2}$ nuclei
used as qubits: $M$ (three hydrogens that share the same chemical environment
and thus appear practically indistinguishable), $C_1$ and $C_2$ (both are carbons)
and (b) graph representation of interactions between qubit nuclei of this molecule.}
\label{ac}
\end{center}
\end{figure}

\end{example}

\begin{example}
Illustrated in Figure \ref{encirc} is a circuit (in terms of NMR
pulses) for the encoding part of the 3-qubit quantum error
correction code taken from \cite{martin}. It consists of a number of
single qubit and two qubit gates. Gates $R_y(90)$ require an RF
pulse of the length proportional to the degree of rotation. Since
the degree of rotation is $90$, $T(R_y(90))=1$. In liquid state NMR
gates $R_z(90)$ and $R_z(-90)$ can be applied through the change of
rotating reference frame and require no delay. Thus,
$T(R_z(90))=T(R_z(-90))=0$. Both $zz$-interactions are $90$-degree,
therefore $T(ZZ(90))=1$.

\begin{figure}[ht]
\begin{center}
\includegraphics[height=24mm]{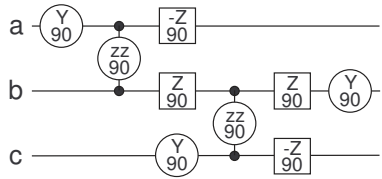}
\caption{Encoding part of the 3-qubit error correcting code.}
\label{encirc}
\end{center}
\end{figure}

\end{example}

\begin{example}
Let us consider the placement problem of the 3-qubit encoding quantum circuit (Figure
\ref{encirc}) into acetyl chloride (Figure \ref{ac}). There are $3!=6$ possible
placement assignments. One possible placement is the following: $a \rightarrow M,\;b \rightarrow C_2$
and $c \rightarrow C_1$. The circuit runtime calculation is illustrated in Table \ref{tab:cc}
for steps $i=1..5$ appearing in columns 2-6 and the first column listing the available qubits.
Single qubit rotations around $Z$ axis are ignored since their contribution to the runtime is zero.
The circuit runtime equals the largest number in the last column, 770. This example
placement is not optimal. The minimal possible value of runtime is equal to $136$, and
is achieved by the placement $a \rightarrow C_2,\;b \rightarrow C_1,\;c \rightarrow M$.

\begin{table}[ht]
\begin{center}
\caption{Cost of $\{a \rightarrow M,\;b \rightarrow C_2,\;c \rightarrow C_1\}$ mapping.}
\begin{tabular}{|c|c|c|c|c|c|}\hline
$time[\;]$ & $Y_a$90 & $ZZ_{ab}$90 & $Y_c$90 & $ZZ_{bc}$90 & $Y_b$90\\ \hline
a       & 8   & 680  & 680 & 680  & 680 \\
b       & 0   & 680  & 680 & 769  & 770 \\
c       & 0   & 0    &  8  & 769  & 769 \\ \hline
\end{tabular}
\label{tab:cc}
\end{center}
\end{table}
\end{example}

\section{Complexity}\label{sec:c}

In this section we show that the placement problem is NP-Complete, and
thus for practical purposes we need to focus on heuristic solutions. In particular, we
consider the problem of finding a Hamiltonian cycle in a graph
$H=(V_H,E_H)$. We next formulate a mapping problem whose solution is
equivalent to the solution of a Hamiltonian cycle problem.

Let the physical environment be a graph with the same set of vertices as
graph $H$, $\{v_1,v_2,...\;,v_m\}$. Edge $(v_i,v_j)$ of the physical environment
has weight one if and only if $(v_i,v_j)$ is not an edge of graph $H$. All
other edges have weight zero. The physical environment, as defined, models the graph $H$
from the Hamiltonian cycle problem. Let the quantum circuit consist of $m$ levels
and be built on $m$ qubits $\{q_1,q_1,...\;,q_m\}$. Let the $i^{\textrm{th}}$
level be composed of a single 2-qubit gate $G(q_i,q_{(i \bmod m)+1})$ with $T(G)=1$.
The runtime of the circuit is thus the sum of the execution runtimes of its gates.
The defined quantum circuit models the Hamiltonian cycle. When the
quantum circuit is mapped into the physical environment such that
its runtime is minimized, the minimal possible value of runtime, which is zero, is
possible to achieve if and only if the physical environment has a
Hamiltonian cycle that goes along the edges with zero weight. This
is equivalent to $H$ having a Hamiltonian cycle.

\section{Heuristic solution}\label{sec:hs}
Before we begin the description of our algorithm, we first make
several observations about the quantum information processing proposals and quantum circuits.
Firstly, in real world experiments, such as \cite{ar:klmn, ar:nmrd}, the qubit-to-qubit
interactions are usually aligned along the chemical bonds, or
otherwise using only the fastest interactions allowed by the
physical environment. Secondly, most often a single qubit gate can
be executed much faster than a two-qubit gate. Third, it may turn
out that a circuit is composed of a number of computations/stages
each of which has its own best mapping into a physical environment.
If the circuit is treated as a whole, no placement will be
efficient because at least one part of such a placement will
be inefficient. Below, we describe an approximate placement
algorithm that uses heuristics derived from these observations.

\subsection{Algorithm}
\noindent {\bf Outline.} The idea of our algorithm is to break a given
circuit into a number of subcircuits such that an efficient placement can
be found for each of the subcircuits. We next connect the individually
placed subcircuits to form the desired computation via swapping/permutation
circuits.

\noindent {\bf Preprocessing.} We first take the physical
environment and establish which interactions are considered the
fastest. One method for doing this is to choose a $Threshold$ such
that if a value of a qubit-to-qubit interaction is below the
$Threshold$, the interaction is considered to be fast, otherwise it
is considered too slow and will not be used. The value of the $Threshold$
may be chosen to be the minimal value such that the graph associated with
fastest interactions is connected, or may be taken directly from the
experimentalists. Either way, we treat the value of the $Threshold$
as a known parameter for our algorithm. Using the $Threshold$ value
ensures that our algorithm comes up with a circuit implementation
that uses only the fastest interactions.

\noindent {\bf Algorithm.}
The algorithm consists of two stages: basic placement and fine tuning,
repeated iteratively. In the basic mapping stage we make sure that in some
part of the circuit all two qubit gates can be
aligned along the fastest interactions of the physical
environment. In the fine tuning stage we consider a given subcircuit for which
a mapping of circuit qubits into the molecule's nuclei along the
fastest interactions exists. We fine tune this matching by shuffling
the solution and taking the actual numbers that represent the length of each gate
(including single qubit gates) into account.

We start by reading in the 2-qubit gates from the circuit into a
workspace $C$ and do it as long as these gates can be arranged along
the fastest interactions provided by the physical environment. As
soon as we reach a gate whose addition will prohibit the alignment
along the fastest interactions, we stop and concentrate on those
gates already in $C$. At this point, we know that the two qubit
gates we have read, $g_1, g_2,...\;,g_s$, can be aligned along the
fastest interactions. Take one of such alignments in the case
when multiple alignments are possible. The alignment itself can be
found by any of the known graph monomorphism techniques, such as
\cite{co:cfsv}. If all monomorphisms can be found reasonably fast
(which is true for all experimentally constructed circuits and
environments we tried due to their small size/suitable structure), we calculate the
circuit runtime for each of them and take the best found. Otherwise,
we apply a simple hill climbing algorithm, where for every qubit
$q_i$ from the circuit such that there exists a two qubit gate $g_j$
among those in $C$ that operates on this qubit, try to map it to any
of $\{v_1,v_2,...\;,v_m\}$ and see if this new placement assignment is better
than the one provided by the initial matching. If it is, change the
way qubit $q_i$ is placed, otherwise, move on to the next qubit. Such
an operation can be repeated until no improvement
can be found or for a set number of iterations.  We refer to this step as fine tuning. After the
matching is fine tuned, we return to the circuit, erase all the
gates from it up to the two qubit gate $g_{s+1}$ that prevented
alignment along the fastest interactions at the previous step and
repeat the above procedure.

The result of the above algorithm is a finite set of finely tuned
placement assignments $P_1, P_2,...\;,P_t$,
but they apply to different subcircuits $C_1,C_2,...\;,C_t$
of the target circuit. The overall computation looks as
follows $C_1E_{1,2}C_2E_{2,3}...$ $E_{t-1,t}C_t$,
where $E_{i,i+1}$ is a circuit composed with SWAP gates such that $E_{i,i+1}$ transforms
the mapping given by $P_i$ into the mapping given by $P_{i+1}$, and initially
all qubits are placed according to $P_1$.

\subsection{Fast permutation circuits}\label{sub:fps}

The algorithm from the previous subsection describes all details of the circuit placement,
except how to construct circuits $E_{i,i+1}$ that would
swap the placement assignment of subcircuit $C_i$ into that for $C_{i+1}$.
In this subsection we discuss how to compose such circuits efficiently
(linear in the number of qubits in the physical environment; and
using only the fastest interactions) and solely in terms of SWAP gates.

\noindent {\bf Problem.} For a given permutation (defined as the permutation transforming the
placement assignment of subcircuit $C_i$ into that for $C_{i+1}$) and a graph defining which
two qubits can be interchanged through a single SWAP (such an {\em adjacency graph} has an edge between
two qubits if it is possible to SWAP their values directly, meaning that
the interaction between the two qubits is considered fast), compose a set of
SWAPs that realize this permutation.

\noindent {\bf Observations.} Note that the interaction graph is connected.

\noindent {\bf Assumptions.} For simplicity, we assume that all SWAP gates applied
to the qubits joined by the edges of adjacency graph $G$ require the same time. Modification
of the algorithm presented below that accounts for the actual costs of SWAPs is
possible, but it is not discussed here. Next, assume that the graph $G$, as well
as its subgraphs, are {\em well separable}.  That is, there exists a constant
$s \in (0,1]$ such that $G$ can be recursively divided into two
connected components $G_1$ and $G_2$, and the ratio of the
number of vertices in the smaller subgraph to the number of
vertices in the larger subgraph is never less than $s$. In the Appendix
we prove that every bounded degree graph with the maximal vertex degree equal $k$
is well separable with parameter $s=\frac{1}{k}$. If we now turn our
attention to the possible quantum architectures, we find that it is not physical to
believe that there exists a scalable quantum architecture that would allow an unbounded number
of direct neighbors. Thus, restricting our attention to bounded degree (and as such,
well separable) graphs does not seem to impose any serious restriction.
The well separability property
certainly holds true for the scalable interaction architectures proposed in the
literature: linear nearest neighbor architecture (1D lattice, $s=1/2$), and 2D
lattices ($s \geq 1/2$). In our experiments, we found that the
interaction graphs for liquid state NMR molecules have $s=1/2$
(which is somewhat better than $s=1/3$ and $s=1/4$ that follow from the
theorem in Appendix).

If a number of pairs with different values is to be swapped, the
runtime of the experimental implementation of such an operation is
associated with the cost of a single SWAP. We make this assumption
because in quantum technologies it is possible (not prohibited, and
often explicitly possible, such as in liquid state NMR) to execute
non-intersecting gates in parallel. Moreover, implementing non-intersecting
gates in parallel is not a luxury, rather a necessity for fault-tolerance
to be possible \cite{co:ab-o}.

\noindent {\bf Goal.} Our target is to minimize the number of logic levels
one needs in order to realize a given permutation as a circuit. Each logic level is composed
with a set of non-intersecting SWAPs operating along the fastest interactions
defined by the adjacency graph.

\noindent {\bf Outline.} We construct the swapping circuit with the help
of a recursive algorithm. At each step our recursion makes sure that certain
quantum information has been propagated to its desired location.

\noindent {\bf Algorithm.} Consider adjacency graph $G$ with $n$ vertices
that shows which SWAPs
can be done directly. Cut this graph into two connected
subgraphs $G_1$ and $G_2$ with the number of vertices equal to or as close
to $n/2$ as possible. To do such a cutting we would
have to cut a few edges, each called a {\em communication channel}.

Consider subgraphs $G_1$ and $G_2$. Color each vertex white if ultimately
(according to the permutation that needs to be realized) we want
to see it in $G_1$ and black if we want to see it in $G_2$. The task is to
permute the colors by swapping colors in the adjacent vertices
such that all white colored vertices go to the $G_1$ part and all
black ones move to $G_2$. The bottleneck is in how many edges
join the two halves $G_1$ and $G_2$---the number of such edges represents the
{\em carrying capacity} of the communication channel. This is because all
white elements from $G_2$ and all black elements from $G_1$ must, at some
point, use one of such communication channels to move to their subgraph.

If all colored vertices can be brought to their subgraph with a linear
cost, $a*n$, then we iterate the algorithm and
have the following figure for the complexity of the entire algorithm:
$C(n) \leq a*n + C(\frac{n}{s+1}),$
where $C(n)$ is the complexity of the algorithm for a size $n$ problem.
The solution for such a recurrence is $C(n)=\frac{a(s+1)n}{s}$.
If the graph has separability factor $s=1/2$, the recurrence
can be rewritten as $C(n) \leq a*n + C(2n/3)$. The latter has solution
\begin{equation}\label{eqn1}
C(n) = 3a*n + const,
\end{equation}
\noindent
providing a linear upper bound for the entire algorithm.

We next discuss how to move all white vertices into the $G_1$ subgraph and,
simultaneously, black ones into the $G_2$ subgraph in linear time $a*n$. Consider a single
subgraph, let us say $G_1$, and the edge(s) of $G$ that we cut to create $G_1$ and $G_2$. First, let us
bring all black vertices as close to the communication channel as
possible. To do so, we suppose that the communication channel consists of a
single edge, otherwise, choose a single edge. We next
cut all loops in $G_1$ such that in the following solution we
will not allow swapping values across all possible edges. After this is done,
we have a rooted tree with the root being the unique vertex that belongs
to $G_1$ and is the end of the communication channel. A rooted tree
has a natural partial order to its vertices, induced by the minimal
length path from a given vertex to the root.

We next apply the following algorithm. Suppose $G_1$ has $k+1$
vertices. At step $i=1..k$, we look at
vertices at depth $k-i$ and their parent at depth $k-i-1$. If
the vertex at depth $k-i$ is black and its parent is white,
we call such a parent a {\em bubble} and interchange it with
the child through applying a SWAP. We also look at all bubbles
introduced on the previous steps. If a bubble has a black child,
we SWAP the two (propagate the bubble). Otherwise, all children of the bubble are white,
and as a result, this vertex is no longer a bubble. Let us note
that once a bubble started ``moving'', it will move each step
of the algorithm until it is no longer a bubble, which happens
only when all its children are white. By the step $i=k$ all
potential bubbles have started moving or have already finished their trip.
It may take an additional $k$ steps for moving bubbles to finish
their movement. By the time step $i=2k$ is completed, every path from a
vertex in $G_1$ to the root will change color at most once (if color
changes it can only change from white to black).

This brings us to the next part:
interchanging white and black nodes in $G_1$ and $G_2$ and actually using the
communication channel. We use the communication channel every odd step in
this part of the algorithm to bring a single black vertex from $G_1$
over to $G_2$ and a single white vertex from $G_2$ to $G_1$. Once
a white vertex arrives in $G_1$ we call it a bubble
and raise it until all its children are white.
A similar operation is performed with graph $G_2$.
Every even step the root vertex of $G_1$ is white (and thus is a bubble)
and one of its children is black (if all children are white, the algorithm
is finished; $G_1$ is guaranteed to contain only white vertices,
and all of them). We propagate the bubble to one of its children to have
a black vertex ready for the next transfer. To get all white nodes to
their side we will need at most $2k$ swaps ($k$ because there cannot be more than
$k$ white nodes missing, and $*2$ because we use communication
channel every second time).

Thus, the total cost of bringing white nodes to $G_1$ and
black ones to $G_2$ is $2k+2k=4k$. Since in the graphs we consider
$s \geq 1/2$, the above expression gives an upper bound of $a=8/3$. Substituting
this into the recurrence solution (\ref{eqn1}) gives the upper bound of
$8n+const$ for the number of levels in a quantum circuit composed
with SWAPs working with the fastest interactions and realizing a given permutation.
By considering the permutation $(n,2,3,...,n-1,1)$ and the chain nearest neighbor
architecture we can show that the above algorithm is asymptotically optimal. \\

\noindent {\bf Physical interpretation.} The
above formulated algorithm can be interpreted in terms of elementary
physics. One can think of a graph as a container with a rather
unusual shape (whatever the shape of the graph is when it is drawn),
black nodes contain liquid, white nodes---air, and edges are tubes
joining the containers. Part $G_1$ is solely on the left and
$G_2$ on the right, so they are well separated in space. In
other words, we have a mixture of air and water in some container.
We next flip the container such that the $G_1$ part comes on top and
$G_2$ is on the bottom.  We then observe how the water falls down
while the air bubbles rise up. Intuition tells us that the water
will flow through the communication channel between $G_1$ and $G_2$
with some constant rate (ignore the pressure, friction and other
physical parameters), so that the water will end up on the bottom in a linear
time. Our algorithm tries to model this situation to some extent,
however, to have an easier mathematical proof of linearity we block
the communication channel for some time. In our implementation of
the above algorithm, we do not block the communication channel in
hopes of achieving a faster solution. We illustrate how this algorithm works with the
following example.

\begin{example}
Consider trans-crotonic acid, which could be used as an environment for up to
a 7-qubit computation. The values of qubits are stored in magnetic spins
of carbon nuclei $C_1,C_2,C_3$ and $C_4$, hydrogen nuclei $H_1$ and $H_2$
and the set of three (practically indistinguishable) hydrogens called $M$.
This molecule is illustrated in Figure \ref{transcrot}. We first cut
the graph of chemical bonds into two maximally balanced graphs in terms of
the equality of the number of nuclei used in the quantum computation.
This is done via cutting across ``cut 1'' (shown in the Figure \ref{transcrot}).
Let the component on the left be called $G_1$ and the component
on the right be called $G_2$. Next, each of the
two connected components is cut into two maximally balanced components. Graph $G_2$
has three vertices and must be cut into two connected graphs. This necessitates
that one of the components will have twice as many vertices as the other,
and thus $s=1/2$. The third and final cutting is trivial.

\begin{figure}[ht]
\begin{center}
\includegraphics[height=33mm]{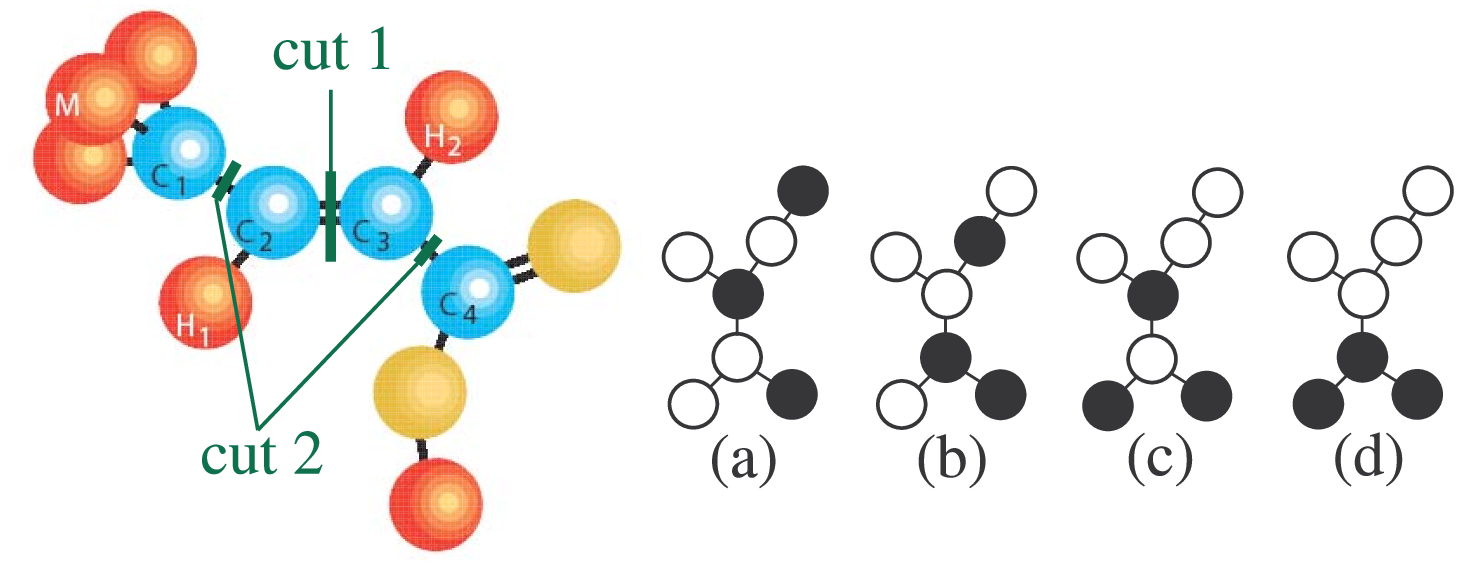}
\caption{Left: Cutting the graph of chemical bonds (fastest interactions) for the
trans-crotonic acid into connected subgraphs. The minimal value of $s$ equals $1/2$.
Right: Permuting values stored in spin-$1/2$ nuclei used in quantum computation.}
\label{transcrot}
\end{center}
\end{figure}

Suppose we want to realize permutation
\begin{displaymath}
\left( \begin{array}{ccccccc}
M & C_1 & H_1 & C_2 & C_3 & H_2 & C_4 \\
C_1  & C_2 & C_3  & C_4 & H_2 & H_1 & M
\end{array} \right).
\end{displaymath}

\noindent Following our water and air metaphor discussed previously, the
vertices contain $Water$--$Air$--$Water$--$Air$--$Air$--$Air$--$Water$.
Once the graph has been cut into the subgraphs as discussed, we
turn the entire graph (i.e. container) $90$ degrees clockwise
(as illustrated in Figure \ref{transcrot}(a)) and observe how the water
falls down. First, the vertices $M$ and $C_1$ and, at the same time, $C_2$ and $C_3$,
exchange their content leading to $Air$--$Water$--$Air$--$Air$--$Water$--$Water$--$Air$,
Figure \ref{transcrot}(b). In the
second step, the contents of $C_1$ and $C_2$ and, at the same, time $C_3$ and $C_4$,
are interchanged leading to $Air$--$Air$--$Air$--$Water$--$Water$--$Air$--$Water$,
Figure \ref{transcrot}(c). Finally, the
contents of $C_2$ and $C_3$ are interchanged leading to
$Air$--$Air$--$Air$--$Air$--$Water$--$Water$--$Water$
(Figure \ref{transcrot}(d)), meaning that all vertices that
were desired in $G_1$ are now moved to $G_1$, and also all vertices
that we wanted to see in $G_2$ are now located in $G_2$.
At this step the problem falls into two completely
separate components that can be handled in parallel:
permute values in a connected graph with 4 vertices and a connected graph with 3 vertices.
\end{example}

\subsection{Implementation and its scalability}
We implemented the above algorithms in C++. Our code consists of two
major parts one of which is the subcircuit placement, and the second
is realization of the swapping algorithm. We next discuss implementation
details of each of these parts as well as how they are used to achieve
the final placement.

Firstly, we use the VFLib graph matching library \cite{www:gm} to solve graph
monomorphism problem which is a basis for the subcircuit placement. The bottleneck
of the entire implementation is the efficiency of computing a solution to the subgraph
monomorphism problem. This is because in order to solve the placement problem
for a circuit with $k$ gates, among which $s$ are two-qubit gates (obviously, $s<k$),
the subgraph monomorphism routine is called at most $2s$ times. All other
operations performed by our algorithm and its implementation are at
most quadratic in $k$.

The efficiency of the VFLib graph matching implementation was
studied in \cite{co:cfsv}. The authors of \cite{co:cfsv} indicate
that the implementation is sufficiently fast for alignment of graphs
containing up to 1000 (and sometimes up to 10,000) nodes. A study of
runtime for aligning a particular type of graphs with around 1000
vertices showed that the alignment took close to one second on average
\cite{co:cfsv}. We thus conclude that our implementation should not
take more than an hour (3600 seconds, in other words, 3600 calls to
the subgraph isomorphism subroutine) to place quantum circuits built on
a few hundred qubits and containing a few thousand gates.

Our basic implementation of the swapping algorithm is exactly as
described in Subsection \ref{sub:fps}, with the communication channel
available for a qubit state transfer at any time. We modified this
algorithm by adding a heuristic called the {\em leaf--target value override}.  This heuristic works by inspecting
leaf values at every stage of the swapping algorithm, and if a desirable value (according to the final
permutation that needs to be realized) can be swapped directly into a
vertex of the physical environment that is a leaf, this operation is
performed. This leaf with the desired value stored
in it is then excluded from the remainder of the swapping algorithm.
The set of leaves of the physical environment graph that still participate
in the algorithm (because they do not yet contain their target value) is
updated dynamically. In practice, this update to the swapping algorithm
helped to reduce the depth of the swapping stage on the order of 0-5\%.

After each mapping is calculated by our monomorphism routine, we greedily select the least cost
mapping.  To improve this greedy solution, we added a depth-2 look ahead algorithm that combines
the cost of a potential mapping with the associated swap cost and all of the potential next stage mappings and swap costs.
The algorithm works as follows.

Given a subcircuit of a maximal size such that it can be placed as a whole we use the VFLib graph monomorphism
routine to come up with monomorphisms $M_1, M_2, \ldots, M_k$ for some fixed value $k$.
For each monomorphism, we construct the swappings $S_1, S_2, \ldots, S_k$ and look for
the next monomorphism, followed by the next swapping stage. At this point
we have examined $k^2$ circuits $M_1S_1M_{1,1}S_{1,1},\; M_1S_1M_{1,2}S_{1,2}, ...,$
$M_1S_1M_{1,k}S_{1,k},\; M_2S_2M_{2,1}S_{2,1},\; M_2S_2M_{2,2}S_{2,2}, ...,$ $M_2S_2M_{2,k}S_{2,k}, $
$..., M_kS_kM_{k,k}S_{k,k}$ (only the current best circuit is actually stored)
with runtime costs $C_{1,1},\; C_{1,2}, ..., C_{k,k}$. We choose a minimal value $C_{i,j}$
corresponding to the partial placement $M_iS_iM_{i,j}S_{i,j}$ and accept
matching $M_i$ followed by the swapping $S_i$. The process is iterated until the
entire circuit is placed. In our implementation, we used $k=100$ to limit the number
of possible monomorphisms considered. Let us note that the number of times a
monomorphism needs to be called is only $2k$ as opposed to the expected $k^2$.
This is because the sets of monomorphisms $\{M_{i,j}| j=1..n\}$ for different values
$i$ are equal.

\section{Experimental results}\label{sec:er}

In the experiments described below we modify the circuit depth calculation to take into account
that it is not necessary to use an existing interaction more than three times to realize any
two-qubit unitary \cite{ar:zvs}. This means that the cost calculation
may be slightly (with the examples we tried, no more than 1\%) lower than it would be
if the above was not applied.

\begin{table*}[t]
\begin{center}
\caption{Mapping experimentally constructed circuits into their physical environment.}
\begin{tabular}{|c|c|c|c|c|c|c|}\hline
\multicolumn{3}{|c|}{Circuit} & \multicolumn{2}{c|}{Environment} & Estimated & Search \\
name & \# gates & \# qubits & name & \# qubits & circuit runtime  &  space size \\ \hline
error correction encoding   & 9 & 3 & acetyl chloride   & 3 & .0136 sec & 6 \\
\cite{martin}, Figure \ref{encirc} &   &   & \cite{martin}, Figure \ref{ac}  &   &   &   \\ \hline
5 bit error correction   & 25 & 5 & trans-crotonic  & 7 & .0779 sec & 2,520 \\
(\cite{ar:klmn}, Fig. 1) &   &   & acid (\cite{ar:klmn}, Fig. 3) &  &  &   \\ \hline
pseudo-cat state  & 54 & 10 & histidine  & 12 & .5170 sec & 239,500,800 \\
preparation (\cite{ar:nmrd}, Fig. 1) &   &   & (\cite{ar:nmrd}, Fig. 2) &   &   &   \\ \hline
\end{tabular}
\label{tab:er1}
\end{center}
\end{table*}

In this section we present three types of experiments. First, we take circuits
that were executed experimentally on an existing hardware. We erase the placement assignment
of the circuit qubits into the physical environment and try to reconstruct it.
Our tool finds the best solutions found by hand by experimentalists (meaning the
tool creates only one workspace and the matching made is equivalent to that by
experimentalists). Table \ref{tab:er1} summarizes the results. The first three columns
list properties of the circuit---a short description of what the circuits does and its
source, the number of gates and qubits in it. The next two columns present properties
of the environment---the name of the environment and its source, and the maximal number of
qubits it supports. The {\em Estimated circuit runtime} column shows a result of the running time
our algorithm formulated---an estimate (actual runtime, depending on the particulars of technology,
may be slightly different to account for the second order effects) of how long it will
take to execute a given circuit. We do not list the actual mapping, because it is equivalent
to that by the experimentalists \cite{ar:klmn, martin, ar:nmrd}. The last column presents
the size of the search space for placing a circuit as a whole (considering $k$ subcircuits
results in power $k$ blow up of the search space size). This tests the quality of our approach via
comparison to the known results and it illustrates that our
tool correctly chooses to consider only a single subcircuit and places it optimally.

We next describe an extended test in which we considered small but potentially
useful computations.   These included the following quantum circuits: phase estimation (a 5-qubit ``phaseest'' circuit), 6-qubit
Quantum Fourier Transformation (QFT) circuit (\cite{bk:nc}, page 219; circuit name ``qft6''),
9-qubit approximate QFT (circuit name ``aqft9''), 10-qubit $[[7,1,3]]$ Steane code X-type
error correction (circuit names ``steane-x/z1'' and ``steane-x/z2'', corresponding to
Fig. 10.16 and 10.17 in \cite{bk:nc}), which, by symmetry and properties of the circuit, can
be thought of as a Z-type error correction, and an approximate 12-qubit QFT (circuit name
``aqft12'').  Also included in the experiment were the existing liquid state NMR molecules with the most qubits whose complete
description we were able to find in the literature. The molecules include a 5-qubit
BOC-($^{13}$C$_2$-$^{15}$N-$^2$D$_2^{\alpha}$-glycine)-fluoride \cite{quant-ph/9905087},
5-qubit pentafluorobutadienyl cyclopentadienyldicarbonyliron \cite{quant-ph/0007017},
7-qubit trans-crotonic acid \cite{ar:klmn}, and 12-qubit histidine \cite{ar:nmrd}.
Note that the latter two molecules are more recent and the experiments with them were done on better equipment.
Thus, it is natural to expect more qubits and a faster associated circuit runtime.
In particular, the experiment with pentafluorobutadienyl cyclopentadienyldicarbonyliron is so ``slow'' that
the choice of parameter $Threshold$ equal 50 or 100 disallows all interactions in the molecule
making the corresponding computation impossible to run (which is indicated with N/A in Table
\ref{tab:er2}).  This test illustrates the importance of using the
swapping stages, studies the relation between the value of $Threshold$ and the total circuit runtime,
shows the quality of the results, and provides a potentially useful data set for experiments.

The results are formulated in Table \ref{tab:er2}. Each entry in this table consists of two
numbers. The first number shows the calculated total runtime of the placed circuit in the given
table row for the $Threshold$ value as declared in the given column, for each physical environment described on top of the given part of the Table.
The second entry of the cell is a natural number in brackets that represents how many
subcircuits our software chooses to do the placement with. Note, that the last column in Table \ref{tab:er2}
lists circuit runtime with the optimal placement when placed without insertion of SWAPs. Every
number in the row smaller than the last number in it is an indication that our swapping
strategy outperformed optimal placement of the circuit as a whole.

In particular, as a part of the analysis of the results in Table \ref{tab:er2} let
us consider the circuit for a 6 qubit QFT and map it into the 7-qubit
trans-crotonic acid molecule.
This circuit is inconvenient for quantum architectures
since it contains a 2-qubit gate for every pair of qubits. It cannot
be executed in a chain nearest neighbor sub-architecture \cite{ar:fdh}
since the longest spin chain in trans-crotonic acid has only five qubits.
As expected, usage of a large $Threshold$ value equal to 10000
results in a single working space, with the runtime of the placed
circuit decreasing when reducing the value of $Threshold$.
However, the number of individually aligned subcircuits grows, and, as a result,
the time spent in swapping the values increases. For a very low $Threshold$
value equal to $50$ (in this case, the adjacency graph of the molecule is disconnected,
which, in fact, is an indication that the value of $Threshold$ is
already too small to give the best placement result), the mapped circuit does too much swapping,
resulting in a high overall cost. The best result is achieved with $Threshold=200$.
This indicates that the {\em quantum circuit placement tool has to use
some rounds of SWAPs to achieve best results}. This is because
the circuit with $Threshold$ value 10000 is placed optimally as whole
(i.e. when no SWAPs are allowed). The best circuit execution time with a
``smart'' multiple-subcircuit placement that we were able to achieve is $.2237$ sec. This is almost twice as fast
as $.4137$ sec, which is the best one can possibly achieve by placing the circuit as a whole. A more dramatic
change in the circuit runtime when considering subcircuits can be observed with the circuits ``phaseest'' and
``steane-x/z''.

\begin{table*}
\begin{center}
\caption{Placement of potentially interesting circuits in the existing physical environments for
different values of the $Threshold$.}
{\small 
 \begin{tabular}{|c|c|c|c|c|c|c|}
	\hline
	\multicolumn{7}{|c|}{Placement with the 5-qubit BOC-($^{13}$C$_2$-$^{15}$N-$^2$D$_2^{\alpha}$-glycine)-fluoride
	molecule \cite{quant-ph/9905087}} \\ \hline
 	Circuit & \multicolumn{6}{|c|}{Threshold} \\
	\hline & 50 & 100 & 200 & 500 & 1000 & 10000 \\
	\hline phaseest & .9980 sec (8) &  .9980 sec (8) & .8167 sec (4) & .8167 sec (4) & .4314 sec (3) & .5632 sec (1) \\
	\hline \hline
	\multicolumn{7}{|c|}{Placement with the 5-qubit pentafluorobutadienyl cyclopentadienyldicarbonyliron molecule \cite{quant-ph/0007017}} \\ \hline
	Circuit & \multicolumn{6}{|c|}{Threshold} \\
	\hline & 50 & 100 & 200 & 500 & 1000 & 10000 \\
	\hline phaseest & N/A & N/A & 8.2092 sec (8) &  7.7179 sec (4) & 7.7179 sec (4) & .3733 sec (1) \\
	\hline \hline
	\multicolumn{7}{|c|}{Placement with the 7-qubit trans-crotonic acid \cite{ar:klmn}} \\ \hline
	Circuit & \multicolumn{6}{|c|}{Threshold} \\
	\hline & 50 & 100 & 200 & 500 & 1000 & 10000 \\
	\hline phaseest & .1636 sec (7) & .0699 sec (4) & .0699 sec (4) & .0700 sec (3) & .2156 sec (2) & .1812 sec (1) \\
	\hline qft6 & .3766 sec (9) & .3294 sec (5) & .2237 sec (5) & .2308 sec (5) & .3120 sec (3) & .4137 sec (1) \\
	\hline \hline
	\multicolumn{7}{|c|}{Placement with the 12-qubit histidine molecule \cite{ar:nmrd}} \\ \hline
 	Circuit & \multicolumn{6}{|c|}{Threshold} \\
	\hline & 50 & 100 & 200 & 500 & 1000 & 10000 \\
	\hline phaseest & 1.2022 sec (7) & .6860 sec (4) & .6860 sec (4) & .1827 sec (3) & .1517 sec (2) & .1870 sec (1) \\
	\hline qft6 & 1.9824 drv (9) & .9519 sec (6) & 1.1607 sec (5) & .3123 sec (4) & .5623 sec (3) & .4412 sec (1) \\	
	\hline aqft9 & 4.3713 sec (15) & 2.5419 sec (10) & 1.3405 sec (8) & 1.5400 sec (7) & 1.4927 sec (4) & 1.3367 sec (1) \\
	\hline steane-x/z1 & 1.7427 sec (10) & 1.1898 sec (4) & 1.3402 sec (4) & 1.6326 sec (4) & .5990 sec (2) & 1.0436 sec (1) \\
	\hline steane-x/z2 & 1.3233 sec (7) & 1.2715 sec (4) & 1.0110 sec (3) & .4166 sec (2) & .4677 sec (2) & .9515 sec (1) \\
	\hline aqft12 & 8.1046 sec (23) & 5.3014 sec (15) & 6.0413 sec (13) & 3.5143 sec (10) & 3.3362 sec (8) & 2.6426 sec (1) \\
	\hline
 \end{tabular}
}
\label{tab:er2}
\end{center}
\end{table*}

In the final experiment, we test the performance and scalability of our implementation.  We begin by generating physical environments for the test with 
$N$ qubits, $x_1, x_2, \ldots , x_N$, that form a linear nearest neighbor architecture.  The interaction time between a pair of neighboring qubits is set 
to the value $.001$ sec per a 90-degree 2-qubit rotation (one can think of it as a 1 KHz quantum processor).  Such an architecture seems very natural 
given the efforts in the experimental implementation of the linear nearest neighbor architectures \cite{ar:mo}.

When constructing the test circuit we first randomly permute qubits into $p_1, p_2, \ldots, p_N$ where $p_j = x_i, 1 \leq j \leq N$, $1 \leq i \leq N$.  
We use this permutation to represent a chain architecture where any qubit $p_j$ is coupled to qubits $p_{j-1}$ and $p_{j+1}$.  We then generate a random 
circuit for execution over this physical environment with $N$ qubits and $N \times \log^2(N)$ gates (we assume gates have maximal ``length'', i.e. 
$T(G)=3$ \cite{ar:zvs}).  We do this by first generating gates by randomly choosing an index $j$ and create a two-qubit gate between $p_j$ and $p_{j-1}$ 
or $p_j$ and $p_{j+1}$, each with probability $1/2$.  After $N \times \log(N)$ gates are created, we again randomly permute our $N$ elements and then 
randomly generate $N \times \log(N)$ gates in the chain nearest neighbor architecture that corresponds to this permutation.  This is repeated $\log(N)$ 
times.  We refer to each of these permutations as ``hidden stages''.  The reason behind constructing such a test circuit is as follows. Some quantum 
computations are likely to consist of a number of fairly short phases that are developed and optimized separately, and then need to be ``glued'' together 
to compose the desired computation. For instance, Shor's factoring algorithm \cite{ar:s} consists of modular exponentiation and quantum Fourier 
transformation circuits (approximate QFT circuits have $O(N \times \log(N))$ gates), each studied as a separate problem in the relevant literature (for 
example, \cite{ar:mi, quant-ph/0703211}); modular exponentiation itself can be broken into a number of simpler arithmetic circuits.

Table \ref{tab:er3} displays the results of this experiment for values of $N = 8$ up to $N = 1024$.  The first column is the number of qubits used in each 
experiment, the second column is the number of gates, and the third is the number of ``hidden stages'', i.e. the number of random permutations used while 
generating the circuit's description.  The fourth column is the number of subcircuits generated by our implementation during placement.  This column 
exactly corresponds to the number of hidden stages, which was expected, as we must have a swapping stage to move our temporary architecture from one 
permutation to the next.  The fifth column is the quantum circuit runtime as computed by our placement algorithm and the final column is the running time 
of our software (user time) to compute the placement. The results in Table \ref{tab:er3} indicate that our implementation can process large (hundreds of 
qubits, tens of thousands gates) circuits in reasonable time\footnote{To illustrate the runtime efficiency of our heuristic-based algorithm we compare the 
runtime of our program implementation to the projected estimated runtime of an exhaustive search procedure that looks through all possible logical to 
physical qubit assignments without using the SWAP gates, when the number of qubits equals 512. Apart from such exhaustive search being unable to come up 
with a meaningful solution (which should be breaking the circuit down into 9 stages), the relevant search runtime is described by a 1167-digit decimal 
number (counting each possible assignment as taking one unit of time). Our heuristic implementation takes a few hours to come up with a sensible 
solution.} and come up with a sensible placement. We used an Athlon 64 X2 3800+ machine with 2 GB RAM memory running Linux 2.6.20 for all our experiments.

\begin{table*}
\begin{center}
\caption{Performance test for circuit placement over chains.}
 \begin{tabular}{|c|c|c|c|c|c|}
	\hline
 	\# of Qubits & \# of Gates & Hidden Stages & \# of Subcircuits & Circuit Runtime & Software Runtime \\
	\hline 8 & 72 & 3 &  3 & .118 sec & 0.02 sec \\
	\hline 16 & 256 & 4 &  4 & .458 sec & 0.12 sec \\
	\hline 32 & 800 & 5 &  5 & .937 sec & 1.34 sec \\
	\hline 64 & 2304 & 6 &  6 & 2.747 sec & 7.52 sec \\
	\hline 128 & 6272 & 7 &  7 & 7.147 sec & 69.63 sec \\
	\hline 256 & 16384 & 8 &  8 & 16.88 sec & 674.96 sec \\
	\hline 512 & 41472 & 9 &  9 & 38.107 sec & 9328.00 sec \\
	\hline 1024 & 102400 & 10 & 10 & 86.2820 sec & 173296.00 sec \\
	\hline
 \end{tabular}
\label{tab:er3}
\end{center}
\end{table*}

\section{Conclusions}\label{sec:conclusions}

In this paper we considered the problem of the efficient assignment of physical qubits
to logical qubits in a given circuit, referred to as placement. We studied theoretical aspects
of the problem and proved that even a simplified place-all-at-once version of the placement problem is NP-Complete.
We thus concentrated on a heuristic solution.
We presented an algorithm for the quantum circuit placement problem and its implementation.
We tested our software on experimental data and computed, within a fraction of a second,
the best results that have been reported by the experimental physicists. Further analysis
indicates that our tool can handle very large quantum circuits
(hundreds of qubits and thousands of gates) in a reasonable time (hours), and it is efficient. We found
that considering subcircuits and swapping their mappings is essential in achieving the best placement results.

Further research may be done in the direction of finding a good balance between the depth of a useful computation and
the depth of the following swapping stage (right now, our method is greedy in that the computational stage is formed
to be as large as possible), and using gate commutation (more generally, circuit identities) to transform an instance
of the circuit placement problem into a possibly more favorable one.

\section*{Acknowledgements}

We wish to acknowledge help of S. Aaronson from the University of Waterloo in proving NP-Completeness
of the simplified version of the circuit mapping problem.  This work was supported by NSERC, DTO-ARO,
CFI, ORDCF, CIFAR, CRC, OIT, and Ontario-MRI.

\section*{Appendix}

\begin{theorem}
Every bounded degree graph of maximal degree $k$ is well separable with parameter $s=\frac{1}{k}$.
\end{theorem}

\begin{proof}
In our proof we start with a small connected subgraph $G_1$ such that the graph $\hat{G_1}$ that complements
it to $G$ is itself connected and iteratively add more vertices to it until at some step $t$ the resulting
graph $G_t$ contains at least $\frac{n-1}{k}$ and at most $n-\frac{n-1}{k}$ vertices, while its complement to
$G$ stays connected.

\begin{figure}[ht]
\begin{center}
\includegraphics[height=45mm]{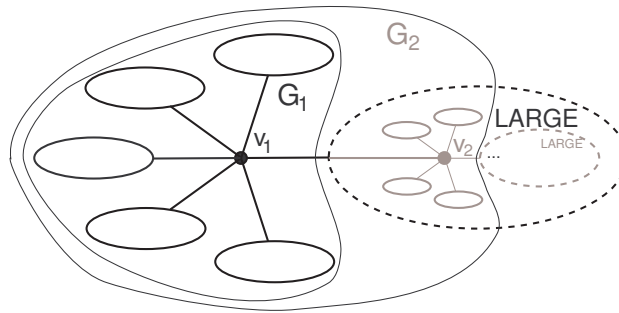}
\caption{Finding a connected component with at least $\frac{n-1}{k}$ and at most $n-\frac{n-1}{k}$ vertices,
whose complement to $G$ is also a connected graph.}
\label{wellsep}
\end{center}
\end{figure}

Consider vertex $v_1 \in G$ whose degree $\deg(v_1) \leq k$.
Cut some edges of $G$ until it becomes a tree, but do so without touching edges connected to $v_1$.
Each of the ellipses illustrated in Figure \ref{wellsep} in black color is now a connected tree.
If one of them, say $i^{\textrm{th}}$, contains at least $\frac{n-1}{k}$ and at most $n-\frac{n-1}{k}$
vertices, cut the edge leading to this subtree and the entire graph breaks into two connected components,
$G_t$ and its complement. At this point, each of the components can be processed iteratively to achieve
well separability. In the case when a component with at least $\frac{n-1}{k}$ and at most
$n-\frac{n-1}{k}$ vertices does not exist, the only possibility is to have one large component having
more than $n-\frac{n-1}{k}$ vertices, and a number of small components. Suppose the edge that joins
$v_1$ and large component is $(v_1,v_2)$. Unite all small components into a graph that we will further
refer to as $G_1$. $G_1$ contains at least $\deg(v_1)$ vertices.

Vertex $v_2$ is connected to $G_1$ and $\deg(v_2)-1$ other connected components. If among other components
there exists one with at least $\frac{n-1}{k}$ and at most $n-\frac{n-1}{k}$ vertices, we have found what we were
looking for. Otherwise, there exists one large (with at least $n-\frac{n-1}{k}$ vertices) component and
a number of small ones. In such case, we join $G_1$ with these components such as shown in Figure \ref{wellsep}
and call this graph $G_2$. Graph $G_2$ has at least one more vertex than $G_1$, because $v_2$ is in $G_2$ but not in $G_1$.
Continue this process with vertices $v_3$, $v_4$, and so on until we accumulated at least $\frac{n-1}{k}$ vertices
in the smaller component or until we found such component otherwise.

It might not be possible to always find a connected component with strictly more than $\frac{n-1}{k}$ and strictly
less than $n-\frac{n-1}{k}$ components, because in the worst case scenario we may have a vertex $v$ of degree
exactly $k$ such that it is connected to pairwise disconnected graphs with exactly $\frac{n-1}{k}$ vertices each. Then, the best we
can do is separate one such component from the remainder of the graph, which will result in the construction of
two connected components with $\frac{n-1}{k}$ and $n-\frac{n-1}{k}$ vertices each. The relation between the number
of vertices in such graphs is
$$\frac{\frac{n-1}{k}}{n-\frac{n-1}{k}}=\frac{n-1}{nk-n+1} \geq \frac{n-1}{nk-k} = \frac{1}{k}.$$

\end{proof}

\end{document}